\documentclass[11pt]{article} 
\usepackage{epsfig} 
\usepackage{amssymb} 
\usepackage[T1]{fontenc}
 
\newcommand{\tr}{\mbox{{\rm tr}}}

\newcommand{\QBit}[1]{{\ensuremath{|#1\rangle}}} 
\newcommand{\tQBit}[1]{{\ensuremath{\langle #1|}}} 
\newcommand{\skQBit}[2]{{\ensuremath{\langle #1|#2\rangle}}}

\newcommand{\mQBit}[2]{{\ensuremath{|#1\rangle\langle #2|}}}

\newcommand{\cA}{{\ensuremath{{\cal A}}}} 
\newcommand{\cB}{{\ensuremath{{\cal B}}}}

\newcommand{\cH}{{\ensuremath{{\cal H}}}}

\newcommand{\cO}{{\ensuremath{{\cal O}}}}

\newcommand{\Prob}{{\mbox{\rm Pr}}}

\newcommand{\NICHT}[1]{}

\newcommand{\binom}[2]{\ensuremath{{#1\choose #2}}}

\newtheorem{Definition}{Definition}

\newtheorem{Theorem}{Theorem} 
\newtheorem{Lemma}{Lemma}

 \newtheorem{Protocol}{Protocol}
\newenvironment{Proof}{\noindent {\em Proof:}  
 \noindent}{\hfill \rule{0.5em}{2ex} } 
\newenvironment{Proofsketch}{\noindent {\em Sketch of the proof:}  
 \noindent}{\hfill \rule{0.5em}{2ex} } 

\title{Using Quantum Oblivious Transfer to Cheat Sensitive Quantum Bit Commitment}
 
\author{ {\large \bf Andreas Jakoby$^1$
                     Maciej Li\'{s}kiewicz$^2$ and 
                     Aleksander M\k{a}dry$^3$}\\[2ex] 
     $^1$ Institut f\"ur Informatik, Universit\"at Frankfurt, Germany \\[1ex]
     $^2$ Institut f\"ur Theoretische Informatik, %\\[0.1ex] 
     Universit\"at zu L\"ubeck,  
     Germany \\[1ex] 
      $^3$ Institute of Computer Science,
     University of Wroc\l{}aw, Poland\\[2ex]} %{\tt a.madry@psz.pl}

\date{}

\begin{document} 
 
\maketitle

\begin{abstract} 
It is well known that unconditionally secure bit commitment 
is impossible even in the quantum world. In this paper a weak variant 
of quantum  bit commitment, introduced independently by 
Aharonov et~al.~\cite{ATVY00} and Hardy and Kent~\cite{HardyKent04}
is investigated. In this variant, the parties require some nonzero 
probability of detecting a cheating, i.e. if Bob, who commits  
a bit $b$ to Alice, changes his mind during the revealing phase
then Alice detects the cheating with a positive probability
(we call this property {\em binding}); and if Alice gains information 
about the committed bit before the revealing phase then Bob discovers 
this with positive probability ({\em sealing}).
In our paper we give quantum  bit commitment scheme that is simultaneously
binding and sealing and we show that if a cheating gives $\varepsilon$ advantage
to a malicious Alice then Bob can detect the cheating with a probability
$\Omega(\varepsilon^2)$. If Bob cheats then Alice's probability 
of detecting the cheating is greater than some fixed constant $\lambda>0$.
This improves the probabilities of cheating detections shown by Hardy and Kent 
and the scheme by  Aharonov et~al.\ who
presented a protocol that is either binding or sealing, but not simultaneously both.

To construct a cheat sensitive quantum  bit commitment scheme 
we use a protocol for a weak quantum 
one-out-of-two oblivious transfer ($\binom{2}{1}$-OT). In this version,
similarly as in the standard definition, Alice has initially 
secret bits $a_0,a_1$ and Bob has a secret selection bit $i$
and if both parties are honest they solve the $\binom{2}{1}$-OT problem
fulfilling the standard security requirements.
However, if Alice is dishonest and she gains some information about the secret 
selection bit then the probability that Bob computes the correct value is 
proportionally small. 
Moreover, if Bob is dishonest and he learns something about both bits,
% but if he does so 
then he is not able to gain full information about one of them.
\end{abstract} 

\section{Introduction}
In bit commitment protocol Bob commits a bit $b$ to Alice in such a way 
that Alice learns nothing (in an information theoretic sense) about $b$ 
during this phase and later on, in the revealing time, Bob cannot change 
his mind. It is well known that unconditionally secure bit commitment is impossible 
even when the parties use quantum communication protocols (\cite{LCh98,Ma97}). 
Thus, much effort has been focused on schemes using some weakened security 
assumptions.

In a weak variant of quantum  bit commitment, introduced independently by 
Aharonov et~al.~\cite{ATVY00} and Hardy and Kent~\cite{HardyKent04}, the 
protocol should guarantee that if one party cheats then the other has good
probability of detecting the mistrustful party. Speaking more precisely, 
we require that if Bob changes his mind during the revealing phase then 
Alice detects the cheating  
with a positive probability (we call this property {\em binding}) 
and if Alice 
learns information about the committed bit before the revealing time then 
Bob discovers the leakage of information 
with positive probability ({\em sealing} property).

In~\cite{HardyKent04} Hardy and Kent give protocol that  
is simultaneously sealing and binding and prove that if Alice (Bob) 
uses a strategy giving $\varepsilon>0$ advantage then Bob (Alice, resp.) 
can detect the cheating
with a probability strictly greater then $0$. The authors do not analyze, 
however, the quantitative dependence of the probability on $\varepsilon$.
In~\cite{ATVY00} Aharonov et~al. present a similar protocol 
to that proposed in~\cite{HardyKent04} such that after depositing phase
either Alice or Bob challenges the other party and 
(1) when Alice asks Bob to reveal $b$ and Bob influences the 
value with advantage $\varepsilon$ then she detects the 
cheating with probability 
$\Omega(\varepsilon^2)$ and 
(2) when Bob challenges Alice to return the depositing 
qubit and Alice predicts $b$ with advantage $\varepsilon$ then Bob
detects the cheating with probability $\Omega(\varepsilon^2)$.
Thus the protocol is either binding or sealing, but not simultaneously 
both (the authors therefore call the protocol a quantum bit escrow). 
Aharonov et~al. left open whether
% the question of achieving 
simultaneous binding and sealing can be achieved.

In our paper we give the first, up to our knowledge, QBC scheme
that is simultaneously binding and sealing such that
if Alice's cheating gives $\varepsilon$ advantage then 
Bob can detect the cheating with a probability which is  
$\Omega(\varepsilon^2)$. If Bob cheats (anyhow) then Alice's probability 
of detecting the cheating is greater than some fixed constant $\lambda>0$, i.e.
when Bob decides to set the value $b$ to $0$ or to $1$ and in the revealing time wants to change his mind
then for any strategy Bob uses the probability that Alice detects 
this attack is greater than $\lambda$.
To construct such scheme we use a protocol for a  weak variant of quantum oblivious 
transfer.

\subsection{Our Contribution}
In the one-out-of-two oblivious transfer problem 
($\binom{2}{1}$-OT, for short) Alice has initially 
two secret bits $a_0,a_1$ and Bob has a secret selection bit $i$.
The aim of a $\binom{2}{1}$-OT protocol 
is disclosing the selected bit $a_i$ to Bob, in such a way
that Bob gains no further information about the other bit
and Alice learns nothing at all. The problem has been proposed by
Even et~al.~\cite{EGL85}, as a generalization of Rabin's
notion for oblivious transfer \cite{Ra81}.
Oblivious transfer is a primitive of central importance
particularly in secure two-party and multi-party computations.
It is well known (\cite{Kilian:Founding:on:OT:1988,BCR86}) that
$\binom{2}{1}$-OT can be used as a basic component to construct protocols 
solving more sophisticated tasks of secure computations
such as two-party oblivious circuit evaluation.
Several secure OT protocols has been proposed in the literature 
\cite {BBCS92,Cr94,CK88} however, even in quantum world, there exists no unconditionally secure 
protocol for $\binom{2}{1}$-OT (see e.g. \cite{Ma97}).  % in information-theoretical sense.

In this paper we define a weak variant of one-out-of-two oblivious transfer.
Similarly as in the standard definition, in a weak $\binom{2}{1}$-OT protocol 
Alice has initially 
secret bits $a_0,a_1$ and Bob has a secret selection bit $i$
and if both parties are honest\footnote{We say that a party is honest if it
never deviate from the given protocol.} 
they solve the $\binom{2}{1}$-OT problem
fulfilling the standard requirements.
However if Alice is dishonest and she gains some information about the secret 
selection bit then the probability that Bob computes the correct value is proportionally decreased. 
Moreover, if Bob is dishonest he can learn about both bits,
but if he does so then he is not able to gain full information about one of them.

In the paper we present a  weak $\binom{2}{1}$-OT protocol
which, speaking informally (precise definitions will be given in Section~\ref{sec:our:def}), 
fulfills the following properties.
\begin{itemize}
\item If both Alice having initially bits $a_0,a_1$ and Bob having bit $i$
      are honest then Bob learns the selected bit $a_i$, but he gains no further 
      information about the other bit and Alice learns nothing.
\item If Bob is honest and has a bit $i$ and Alice learns $i$ 
      with advantage $\varepsilon$ then for all $a_0, a_1\in \{0,1\}$ the probability 
      that Bob computes the correct value $a_i$, when the protocol completes, is 
      at most $1-\Omega(\varepsilon^2)$.
\item If Alice is honest and has bits $a_0,a_1$ 
      then for every $i\in\{0,1\}$ it is true that 
      if Bob can predict the value $a_{1-i}$ with advantage $\varepsilon$
      then the probability that Bob learns correctly $a_i$ is at most
      $1-\Omega(\varepsilon^2)$.
\end{itemize}
The protocol can be used e.g.\ by the mistrustful parties for which 
computing the correct result of $\binom{2}{1}$-OT is much more preferential 
than gaining addition information.
In this paper we show an application of the protocol for  
parties who require some nonzero probability of detecting a cheating.   
Let us consider the following bit commitment protocol, where 
$v:=OT((a_0, a_1), i)$ means, for short, that Alice having initially 
$a_0, a_1$ and Bob knowing $i$ perform the weak $\binom{2}{1}$-OT protocol  
and when the protocol completes Bob knows the result $v$.
\begin{Protocol}[Cheat sensitive QBC] 
\label{Protocol:weakQBC} $B$ commits bit $b$;
 \vspace*{-3mm}
 \begin{itemize}
 \item {\bf Depositing phase}\\
  %\begin{itemize}
  %  \item[1.] 
1. $A$ chooses randomly bits $a_0, a_1, a_2, a_3$;
              $B$ chooses randomly bits $b'$ and $c$;\\
  %  \item[2.] 
2. $A$ and $B$ compute\\
               \hspace*{5mm}$v_0:=OT((a_0, a_1), b');\ v_1:=OT((a_2, a_3), b)$ if $c=0$ or \\
               \hspace*{5mm}$v_0:=OT((a_0, a_1), b);\ v_1:=OT((a_2, a_3), b')$ if $c=1$.\\
  %  \item[3.] 
3. $B$ reveals $c$.
% \end{itemize}
 \item {\bf Revealing phase} $B$ reveals $b$;\\
%   \begin{itemize}
%     \item 
$\circ$    Sealing test: $A$ sends to $B$ $a_{2c}, a_{2c+1}$;
           $B$ rejects when $v_c\neq OT((a_{2c}, a_{2c+1}), b')$. \\
%     \item 
$\circ$ Binding test: $B$ sends to $A$  $v_{1-c}$;
           $A$ rejects when  $v_{1-c}\neq OT((a_{2-2c)}, a_{3-2c}), b)$.
%  \end{itemize}
 \end{itemize}
%\end{tabular}
\end{Protocol}
One of the main results of this paper says that using our weak $\binom{2}{1}$-OT
protocol, the bit commitment protocol above has the following properties:
(1) If both Alice and Bob are honest, then before revealing time Alice 
gains no information about $b$ and at the revealing phase both Bob and Alice accept;
(2) if Alice learns $b$ with advantage $\varepsilon$ then Bob
detects cheating with probability $\Omega(\varepsilon^2)$, and 
(3) if Bob tries to change $b$ during the revealing phase 
then for any strategy he uses the probability that Alice detects 
the cheating is greater than some positive constant.

The paper is organized as follows. In Section \ref{sec:prel} some basic 
quantum preliminaries are given. In Section \ref{sec:our:def} we define 
formally properties of a weak $\binom{2}{1}$-OT protocol and prove that 
the given scheme fulfills the properties. Section \ref{sec:qot}
gives formal definition of binding and sealing and proves that 
Protocol~\ref{Protocol:weakQBC} is simultaneously
binding and sealing.

\section{Preliminaries}\label{sec:prel}

The model of two-party computation we use in this paper is essentially the 
same as defined in \cite{ATVY00}. We assume that the reader is already 
familiar with basics of quantum cryptography (see \cite{ATVY00} for an 
exemplary summary of results that will be used in the following).

Let $\QBit{0}$,$\QBit{1}$ be an encoding of classical bits in our computational (perpendicular) basis. Let $\QBit{0_{\times}}=\frac{1}{\sqrt{2}}(\QBit{0}-\QBit{1})$, $\QBit{1_{\times}}=\frac{1}{\sqrt{2}}(\QBit{0}+\QBit{1})$ be an encoding of classical bits in diagonal basis. By $R_\alpha$, $\alpha\in\{0,\frac{1}{2}, 1\}$, we denote the unitary operation of rotation by an angle of $\alpha\cdot\pi/2$. 
More formally:
$$
R_\alpha :=  
\left(
\begin{array}[c]{rr}
   \cos (\alpha\cdot \frac{\pi}{2}) & \sin (\alpha\cdot \frac{\pi}{2})\\
 - \sin (\alpha\cdot \frac{\pi}{2}) & \cos (\alpha\cdot \frac{\pi}{2})
\end{array}\right)\ 
$$

We should note that this operation allows us to exchange between the bit encoding in perpendicular and in diagonal basis. Moreover, by applying $R_{1}$ we can flip the value of the bit encoded in any of those two bases.   

For a mixed quantum state $\rho$ and a measurement $\cO$
on $\rho$, let $\rho^{\cO}$ denote the classical distribution
on the possible results obtained by measuring $\rho$ according to 
$\cO$, i.e. $\rho^{\cO}$ is some distribution 
$p_1,\ldots,p_t$ where $p_i$ denotes the probability
that we get result $i$. 
%
%\begin{Definition}\label{Def:trace:prob:dist}
We use $L_1$-norm to measure distance between two probability distributions
$p=(p_1,\ldots,p_t)$ and $q=(q_1,\ldots,q_t)$
over $\{1,2,\ldots,t\}$:\
$
  |p-q|_1 = \frac{1}{2}\sum_{i=1}^t |p_i - q_i|.
$ 
%\end{Definition}

Let $||A||_t = \tr(\sqrt{A^{\dagger}A})$, where $\tr(A)$ denotes trace of matrix $A$.
A fundamental theorem gives us a bound on $L_1$-norm for the probability 
distributions on the measurement results:

\begin{Theorem}[see \cite{AKN98}] \label{Th:AKN}
Let $\rho_0$, $\rho_1$ be two density matrices on the
same Hilbert space $\cH$. Then for any generalized measurement $\cO$\
$
  |\rho_0^{\cO} - \rho_1^{\cO}|_1 \le \frac{1}{2}||\rho_0 - \rho_1||_t.
$
% Furthermore, the 
This bound is tight and the orthogonal measurement
$\cO$ that projects a state on the eigenvectors of $\rho_0 - \rho_1$
achieves it.
\end{Theorem}
A well-known result states that if $\QBit{\phi_1}$, $\QBit{\phi_2}$ are pure states, then 
$
%\begin{equation}  \label{eq:trace:pure}
  ||\ \QBit{\phi_1}\tQBit{\phi_1} - \QBit{\phi_2}\tQBit{\phi_2}\ ||_t =
  2\sqrt{1-|\skQBit{\phi_1}{\phi_2}|^2}.
%\end{equation}
$

\begin{Lemma} \label{Lemm:AB:prob:trace} 
% {\bf ToDo: for more general setting.}\\  
Suppose Bob has a bit $b$ s.t. $\Prob[b=0]=1/2$
and let Alice generate a state with two quantum registers. 
Assume she sends the second register to Bob, then 
Bob depending on $b$ makes some transformation on his part 
and sends the result back to Alice. Denote by $\rho_0$ 
density matrix of the resulting state for $b=0$ 
and by $\rho_1$ density matrix of the state for $b=1$. 
Then for any measurement $\cO$  Alice makes 
and a value $v$ Alice learns we have 
$
  \Prob_{b\in_R \{0,1\}}[v=b] \ \le\ 1/2 + \frac{|\rho_0^{\cO} - \rho_1^{\cO}|_1}{2}.
$
\end{Lemma}

The proof of this lemma follows by some straight forward
calculations and will be skipped in this extended abstract.
We will use some obvious variations of this lemma to bound
the advantage of Alice resp. Bob in what will follow.

\section{Weak Oblivious Transfer}
\label{sec:our:def}

In this section we give the formal definition of the weak $\binom{2}{1}$-OT
protocol and then present protocol for this problem.

\begin{Definition}
We say that a two-party quantum protocol between Alice and Bob is a
$(\delta, \varepsilon)$-weak $\binom{2}{1}$-OT protocol if the following 
requirements hold.
\begin{itemize}
\item If both Alice depositing initially bits $a_0,a_1$ and Bob having bit $i$
      are honest then Bob learns the selected bit $a_i$  but in such a way
      that he gains no further information about the other bit
      and Alice learns nothing.
\item Whenever Bob is honest and has a selection bit $i$, with $\Prob[i=0]=1/2$,
      then for every strategy used by Alice, every value $i'$ Alice learns about $i$
      and for any value $a'$  Bob learns at the end of the computation
      it holds that for all $a_0, a_1\in \{0,1\}$ 
      \[
         \mbox{if\ $\Prob_{i\in_R\{0,1\}}[i'=i]\ge 1/2 + \delta$
         \  then\  $\Prob_{i\in_R\{0,1\}}[a'=a_{i}]\le 1-\varepsilon$.}
       \]
\item Whenever Alice is honest and deposits bits $a_0,a_1$, with $\Prob[a_i=0]=1/2$,
      then for every strategy used by Bob, all values $a'_0, a'_1$ Bob learns about 
      $a_0, a_1$, resp. it holds that for all $i\in \{0,1\}$ %\\
      %\centerline{
         $\mbox{if\ $\Prob_{a_0,a_1\in_R\{0,1\}}[a'_{1-i}=a_{1-i}]\ge 1/2 + \delta$
          \ then\  $\Prob_{a_0,a_1\in_R\{0,1\}}[\mbox{$a'_i = a_i$}]\le 1-\varepsilon$.}$
      % }
\end{itemize}
\end{Definition}

\begin{Protocol}[$\binom{2}{1}$-OT function] \label{Protocol:OT}
             {\em Input} $A: a_0,a_1\in \{0,1\}, B: i \in \{0,1\}$;
             {\em Output} $B: a_i$.\\% \\[2mm]
%\begin{tabular}{ll}
% \vspace*{-3mm}
% \begin{itemize}
%\item[1.] 
1.  $A$ chooses randomly
  $\alpha\in_R\{0,\frac{1}{2}\}$ and $h\in_R\{0,1\}$ and sends to $B$: \\ % \\[2mm]
  \centerline{$R_\alpha\QBit{a_1\oplus h}\otimes R_\alpha\QBit{a_0\oplus h}$}\\
%\item[2.] 
2.  $B$ receives $\QBit{\Phi_1}\otimes\QBit{\Phi_0}$,
  chooses randomly $\beta\in_R\{0,1\}$
  and sends
  $R_\beta\QBit{\Phi_i}$ back  to $A$.\\
%\item[3.] 
3.  $A$ receives $\QBit{\Phi}$, computes $R_\alpha^{-1}\QBit{\Phi}$,
  measures the state in computational basis obtaining the\\
\hspace*{2mm} result $n$
  and sends $m=n\oplus h$ to $B$.\\
%\item[4.] 
4.        $B$ receives $m$ and computes $a_{i}=m\oplus \beta$.
% \end{itemize}
%\end{tabular}
\end{Protocol}

Here, as usually, $\otimes$ denotes xor. 
Note that this protocol computes ${2\choose 1}$-OT correctly if both
parties are honest.
We will now focus on the question whether 
Protocol \ref{Protocol:OT} still retains 
security if we use it against malicious parties.
The following theorem follows from Lemma~\ref{l:mal:Al} and~\ref{lem:mal:Bob}
which will be proven in the remaining part of this section:

\begin{Theorem} \label{Th:CQS:OT}
Protocol~\ref{Protocol:OT} is
$(O(\sqrt[2]{\varepsilon}),\varepsilon)$-weak  $\binom{2}{1}$-OT protocol.
%$(O(\sqrt[2]{\varepsilon}),\varepsilon)$-secure against malicious parties.
%More precisely it is $(4\sqrt[2]{\varepsilon}),\varepsilon)$-secure against 
%malicious Alice and $(16\sqrt[2]{2\varepsilon},\varepsilon)$-secure against 
%malicious Bob.
\end{Theorem}

%The Theorem follows from Lemma~\ref{l:mal:Al} and \ref{lem:mal:Bob}.

\subsection{Malicious Alice}

\begin{Lemma} \label{l:mal:Al}
      Let Alice and Bob perform Protocol \ref{Protocol:OT} and
      assume Bob is honest and deposits a bit $i$, with $\Prob[i=0]=1/2$.
      Then for every strategy used by Alice, every value $i'$ Alice learns about $i$
      and for any value $a'$  Bob learns at the end of the computation
      it holds that for all $a_0, a_1\in \{0,1\}$
      \(
         \mbox{if\ $\Prob_{i\in_R\{0,1\}}[a'=a_{i}]\ge 1-\varepsilon$
         \  then\  $\Prob_{i\in_R\{0,1\}}[i'=i]\le 1/2 + 16\sqrt{\varepsilon}$.}
       \)
%Protocol~\ref{Protocol:OT} is $(4\sqrt[2]{2\varepsilon},\varepsilon)$-secure with respect to Alice.
\end{Lemma}
\begin{Proof}
                Any cheating strategy $\cA$ of Alice can be described as preparing some state 
$\QBit{\Phi}=\sum_{x\in \{0,1\}^2} \QBit{v_{x},x}$, sending the two rightmost qubits to Bob and 
perform some measurement $\{H_{0},H_{1},H_{2},H_{3}\}$ on this what she gets back after Bob's round, 
where $H_{0}$,$H_{1}$,$H_{2}$, $H_3$ are four pairwise orthogonal subspaces being a division of whole 
Hilbert space that comes into play, such that, for $l,k=0,1$, if our measurement indicates the 
outcome corresponding to $H_{2k+l}$ then it reflects Alice's belief that $i=l$ and that the message 
$m=k$ should be sent to Bob.

%Let $\sqrt{1-\delta}=||\QBit{\Phi_{S}}||$.
%By $\QBit{\Phi_{a,b}}$ we denote the state $\QBit{\Phi}$ (it includes the part being
%held by Bob) after round 2 in case when $i=a$ and $\beta=b$. In our further considerations
%we allow Alice to see the whole $\QBit{\Phi_{a,b}}$ which can only increase her gain,
%but Bob has to xor both qubits with $\beta$ (it doesn't change anything since in
%original protocol Alice doesn't receive the second qubit back).
%
%To get an intuitive insight into that what will follow now, we should notice that Alice
%in order to gain knowledge about $i$ has to generate a state $\QBit{\Phi}$ having
%high content of antysymmetric part.
%
%        Let $\QBit{\Phi_{S}}=\QBit{v_{00},00}+\QBit{v_{11},11}$,
%$\QBit{\Phi_{A}}=\QBit{v_{01},01}+\QBit{v_{10},10}$. That is,
%$\QBit{\Phi_{S}}$ is a part of the state  that is symmetric with respect to qubits
%being sent to Bob and $\QBit{\Phi_{A}}$ is the rest being anti-symmetric.
%
%        Assume now, that $a_{0}\oplus a_{1}=0$. Let $\sqrt{1-\delta}=||\QBit{\Phi_{S}}||$.
%Now, by $\QBit{\Phi_{a,b}}$ we denote the state $\QBit{\Phi}$ (it includes the part being
%held by Bob) after round 2 in case when $i=a$ and $\beta=b$. In our further calculation we
%allow Alice to see the whole $\QBit{\Phi_{a,b}}$ which can only increase her gain, but Bob
%has to xor both qubits with $\beta$ (it doesn't change anything since in original protocol
%Alice doesn't receive the second qubit back).

Assume now, that $a_{0}\oplus a_{1}=0$. We should note that in this case $m\oplus a_{0}=\beta$. 
So Alice, in order to ensure the correct result of the protocol, 
has to indicate the value of $\beta$. Let $\QBit{S}=\QBit{v_{00},00}+\QBit{v_{11},11}$, 
$\QBit{A}=\QBit{v_{01},01}+\QBit{v_{10},10}$. That is, $\QBit{S}$ is a part of the state  
that is symmetric with respect to qubits being sent to Bob and $\QBit{A}$ is the rest being 
anti-symmetric.
%> And let $||\QBit{S}||=\sqrt{1-\delta}$.

Let $\rho_{a,b}$ be a density matrix of Alice's system after Bob's round, corresponding
to $i=a$ and $\beta=b$. After some calculations we get:

\[
\begin{array}{rcl}
\rho_{0,0}&=& \sum_{x=(x_1,x_2)\in\{0,1\}^2} \mQBit{v_{x}x_1}{v_{x}x_1} \\
& & +
                 \mQBit{v_{00}0}{v_{10}1}+\mQBit{v_{10}1}{v_{00}0}+ \mQBit{v_{11}1}{v_{01}0}+
                 \mQBit{v_{01}0}{v_{11}1}\\
\rho_{0,1}&=& \sum_{x=(x_1,x_2)\in\{0,1\}^2} \mQBit{v_{x}\overline{x_1}}{v_{x}\overline{x_1}}\\
& & -

\mQBit{v_{00}1}{v_{10}0}-\mQBit{v_{10}0}{v_{00}1}-\mQBit{v_{11}0}{v_{01}1}-
                 \mQBit{v_{01}1}{v_{11}0}\\
\rho_{1,0}&=& \sum_{x=(x_1,x_2)\in\{0,1\}^2} \mQBit{v_{x}x_2}{v_{x}x_2} \\
   & &   + \mQBit{v_{00}0}{v_{01}1} +
         \mQBit{v_{01}1}{v_{00}0}+ \mQBit{v_{11}1}{v_{10}0}+\mQBit{v_{10}0}{v_{11}1}\\
\rho_{1,1}&=& \sum_{x=(x_1,x_2)\in\{0,1\}^2} \mQBit{v_{x}\overline{x_2}}{v_{x}\overline{x_2}}\\
& & -

\mQBit{v_{00}1}{v_{01}0}-\mQBit{v_{01}0}{v_{00}1}-\mQBit{v_{11}0}{v_{10}1}-
                 \mQBit{v_{10}1}{v_{11}0}\ .
\end{array}
\]
where $\overline{x_t}$ means flipping bit $x_t$, i.e. $\overline{x_t}=1-x_t$.

We look first onto possibilities of Alice's dishonest behaviour. In order to cheat, Alice
has to distinguish between density matrices $\gamma_l=\frac{1}{2}\rho_{l,0}+\frac{1}{2}\rho_{l,1}$,
where $\gamma_l$ corresponds to $i=l$. By examination of the difference of those matrices we get after
some calculations that:
$$
\gamma_0-\gamma_1= \frac{1}{2}\mQBit{V_S0}{V_A1}+\frac{1}{2}\mQBit{V_A1}{V_S0}-
                \frac{1}{2}\mQBit{V_S1}{V_A0}-\frac{1}{2}\mQBit{V_A0}{V_S1}
$$
where $\QBit{V_S}=\QBit{v_{00}}+\QBit{v_{11}}$ and $\QBit{V_A}=\QBit{v_{10}}-\QBit{v_{01}}$.
We can easily adapt Lemma~\ref{Lemm:AB:prob:trace} to show that
the advantage $\delta$ of Alice is at most  
$\sum_{l=0}^3\sigma_l$
% 0 +\sigma_1\le 2\cdot\max\{\sigma_0,\sigma_1\}$
where
\[
\begin{array}{rcl}
\sigma_l \ = \ |tr(H_l(\gamma_0-\gamma_1){H_l}^\dagger)|  
& \le & 
\sum_{j\in\{0,1\}} \frac{1}{2}|tr (H_l(\mQBit{V_S(j-1)}{V_Aj}+\mQBit{V_Aj}{V_S(j-1)} ){H_l}^\dagger)|\\[2mm]
%%& \le & 
%%\sum_{j\in\{0,1\}}
%% |\Re(\skQBit{O_{j}^{l}}{H_l V_Aj}\skQBit{V_S(1-j){H_l}^\dagger}{O_{j}^{l}})|\\[2mm]
& \le & 
\sum_{j\in\{0,1\}}(|\skQBit{O_{j}^{l}}{V_Aj}|\cdot |\skQBit{V_S(1-j)}{O_{j}^{l}}|)\\[2mm]
&\leq& \sum_{j\in\{0,1\}}|\skQBit{O_{j}^{l}}{V_Aj}|
%%\sum_{j\in\{0,1\}}   |\skQBit{V_S (j-1)}{{H_l}^\dagger H_l V_A j}|\\[2mm]
%%& \le & \sum_{j\in\{0,1\}}   \sqrt{|\skQBit{V_S (j-1)}{H_l V_S (j-1)}|}   
%%                            \sqrt{|\skQBit{V_A j}{H_l V_A j}|} \\[2mm]
%%& \le & \sum_{j\in\{0,1\}}|\skQBit{O_{j}^{l}}{V_Aj}|\ ,
\end{array}
\]
% (maybe some comment here about obtaining this inequality..)
and $\QBit{O_{j}^{l}}$ is an orthogonal, normalized projection of $\QBit{V_Aj}$
onto subspace $H_{l}$.
The second inequality is true because we have  
$tr(H_l \mQBit{V_Aj}{\psi}{H_l}^\dag)= \skQBit{O_{j}^{l}}{V_Aj} \skQBit{\psi}{O_{j}^{l}}$
for every state  $\QBit{\psi}$.
%%To get that the second inequality is true, we evaluateuse, 
%%for $j\in \{0,1\}$,
%%trace of the corresponding matrix by definition using orthonormal 
%%basis including $\QBit{O_{j}^{l}}$ as the first element. 
%%To see that the third inequality is true we take 
%%$H_l = \mQBit{O_{j}^{l}}{O_{j}^{l}} +\sum_{t_j}  \mQBit{t_j}{t_j}$,
%%where $\QBit{t_j}$ is an extension of an orhonormal basis of $H_l$ including 
%% $\QBit{O_{j}^{l}}$ as the first element. 

Let $j_l$ be the index for which
$|\skQBit{O_{j_l}^{l}}{V_Aj_l}|\geq |\skQBit{O_{1-j_l}^{l}}{V_A(1-j_l)}|$.
Clearly, $\sigma_l\leq 2|\skQBit{O_{j_l}^{l}}{V_Aj_l}|$.
Moreover, we assume that $\sigma_{0}+\sigma_{1}\geq \sigma_{2}+\sigma_{3}$.
If this is not the case we could satisfy this condition by altering the strategy $\cA$ of Alice (by appropriate rotation of her basis) in such a way that
% by adding a 
% local transformation to $\cA$, 
% being flipping of both qubits received from Bob - 
% this would 
 the definitions of $H_k$ and $H_{k+2}$ would swap leaving everything else unchanged.  

We look now on the probability of obtaining the correct result by Alice. The probability $p_{0}$ of
Alice getting outcome $\beta=0$ in case of $\beta=1$ is at least
\[
\begin{array}{l}
p_{0}\geq \frac{1}{2}\tQBit{O_{j_0}^{0}}\rho_{0,1}\QBit{O_{j_0}^{0}}+

\frac{1}{2}\tQBit{O_{j_0}^{0}}\rho_{1,1}\QBit{O_{j_0}^{0}}=\\[1mm]
%\frac{1}{2}|\skQBit{O_{j_0}^{0}}{v_{01}0}|^2+\frac{1}{2}|\skQBit{O_{j_0}^{0}}{v_{01}1}|^2+\frac{1}{2}|\skQBit{O_{j_0}^{0}}{v_{10}0}|^2+\frac{1}{2}|\skQBit{O_{j_0}^{0}}{v_{10}1}|^2+\\
\quad\frac{1}{2}|\skQBit{O_{j_0}^{0}}{v_{00}1}-\skQBit{O_{j_0}^{0}}{v_{01}0}|^2+

\frac{1}{2}|\skQBit{O_{j_0}^{0}}{v_{00}1}-\skQBit{O_{j_0}^{0}}{v_{10}0}|^2\\[1mm]
\qquad+\frac{1}{2}|\skQBit{O_{j_0}^{0}}{v_{11}0}-\skQBit{O_{j_0}^{0}}{v_{01}1}|^2+

\frac{1}{2}|\skQBit{O_{j_0}^{0}}{v_{11}0}-\skQBit{O_{j_0}^{0}}{v_{10}1}|^2\ .
\end{array}
\]

So, by inequality $|a-b|^2+|a-c|^2\geq \frac{1}{2}|b-c|^2$ we get that
\[
\begin{array}{l}
p_{0} \ \geq \ \frac{1}{4}|\skQBit{O_{j_0}^{0}}{v_{01}0}-
\skQBit{O_{j_0}^{0}}{v_{10}0}|^2+\frac{1}{4}|\skQBit{O_{j_0}^{0}}{v_{01}1}-
\skQBit{O_{j_0}^{0}}{v_{10}1}|^2 \\[1mm]
\quad
 = \ \frac{1}{4}|\skQBit{O_{j_0}^{0}}{V_A0}|^2+
\frac{1}{4}|\skQBit{O_{j_0}^{0}}{V_A1}|^2 \ \geq \ \frac{1}{16}\sigma_0^2.
\end{array}
\]
Similar calculation of the probability $p_{1}$ of getting outcome $\beta=1$ in case of $\beta=0$
yields that the probability of computing wrong result is at least
$$
\Prob[\beta'\neq \beta]=\Prob[\beta\oplus m\neq a_i]\geq 
\frac{1}{16}(\sigma_0^2+\sigma_1^2)\geq \frac{1}{256} (\sum_{l=0}^{3}\sigma_l)^2.
$$
Hence, the lemma holds for the case $a_0\oplus a_1=0$.

Since in case of $a_0\oplus a_1=1$ the reasoning is completely analogous - we exchange only the
roles of $\QBit{V_S}$ and $\QBit{V_A}$ and Alice has to know the value of $\beta\oplus i$
in order to give the correct answer to Bob,
the proof is concluded.
\end{Proof}
%[Zle oznaczenia !!!]

To see that quadratical bound imposed by the above lemma can be met, consider $\QBit{\Phi}=\sqrt{1-\varepsilon}\QBit{000}+\sqrt{\varepsilon}\QBit{110}$. Intuitively, we label the symmetric and anti-symmetric part of $\QBit{\Phi}$ with $0$ and $1$. Let $H_{2}=\mQBit{01}{01}$, $H_3=0$. One can easily calculate that
$$
\rho_{0,0}= (1-\varepsilon) \mQBit{00}{00}+\sqrt{\varepsilon(1-\varepsilon)}(\mQBit{00}{11}+\mQBit{11}{00})+\varepsilon\mQBit{11}{11}
$$
$$
\rho_{1,0}= (1-\varepsilon) \mQBit{00}{00}+\varepsilon\mQBit{10}{10}  
$$
and therefore $||\rho_{0,0}-\rho_{1,0}||_t\geq \sqrt{\varepsilon(1-\varepsilon)}- 2\varepsilon$. So, by Theoren~\ref{Th:AKN} 
there exists a measurement $\{H_{0}, H_{1}\}$ allowing us to distinguish between those two density matrices with $\sqrt{\varepsilon(1-\varepsilon)}- 2\varepsilon$ accuracy and moreover $H_{2},H_{3}\bot H_{0},H_{1}$ since $tr(H_{2}\rho_{0,0}H_{2}^\dagger)=tr(H_{2}\rho_{1,0}H_{2}^\dagger)=0$. Now, let $M=\{H_{0},H_{1},H_{2},H_{3}\}$ be Alice's measurement. To cheat, we use the following strategy $\cA$ corresponding to her input $a_0=a_1=0$. Alice sends $\QBit{\Phi}$ to Bob, after receiving the qubit back she applies the measurement $M$. If the outcome is $H_{2}$ then she answers $a_{0}\oplus \beta=1$ to Bob and sets $i'=0$ with probability $\frac{1}{2}$, in the other case she sends $a_{0}\oplus \beta=0$ to Bob and according to the outcome being $0$ or $1$ she sets $i'=0$ ($i'=1$).

To see that this strategy gives correct result with probability greater than $1-\varepsilon$ we should note that probability of outcome $H_{2}$ in case of $\beta=0$ is $0$ and in case of $\beta=1$ is $1-\varepsilon$. Therefore, since $\beta=0$ with probability $\frac{1}{2}$, our advantage in determining the input of Bob is greater than $\frac{1}{2}\sqrt{\varepsilon}-\frac{3}{2}\varepsilon$.

\subsection{Malicious Bob}

Now, we analyze Bob's possibility of cheating.

        \begin{Lemma} \label{lem:mal:Bob}
      Let Alice and Bob perform Protocol \ref{Protocol:OT}.
      Assume Alice is honest and deposits bits $a_0,a_1$, with $\Prob[a_i=0]=1/2$.
      Then for every strategy used by Bob and all values $a'_0, a'_1$ which Bob learns about 
      $a_0, a_1$, it holds that: for all $i\in \{0,1\}$ 
\[
  \mbox{if\ $\Prob_{a_0,a_1\in_R\{0,1\}}[\mbox{$a'_i = a_i$}]\ge 1-\varepsilon^2$
  \ then\  $\Prob_{a_0,a_1\in_R\{0,1\}}[a'_{1-i}=a_{1-i}]\le 1/2 + 16\sqrt{2}\varepsilon$.}
\]     
% Protocol~\ref{Protocol:OT} is $(16\sqrt{2\varepsilon},\varepsilon)$-secure 
%with respect to Bob.
       \end{Lemma}
        \begin{Proof}
Consider some malicious strategy $\cB$ of Bob. Wlog we may assume that
the probability of $a'_0 = a_0$ is greater than the probability of $a'_1 = a_1$.
Our aim is to show that 
\[
 \mbox{if\ $\Prob_{a_0,a_1\in_R\{0,1\}}[\mbox{$a'_0 \neq a_0$}]\le \varepsilon^2$
       \ then\  $\Prob_{a_0,a_1\in_R\{0,1\}}[a'_{1}=a_{1}]\le 1/2 + 16\sqrt{2}\varepsilon$.}
\]

Strategy $\cB$ can be think of as a two step process. First a unitary transformation 
$U$ is acting on  $\QBit{\Phi_{a_0,a_1,h}}=
  \QBit{v}\otimes R_\alpha\QBit{a_1\oplus h}\otimes R_\alpha\QBit{a_0\oplus h}$, 
where $v$ is an ancillary state\footnote{Note that this does not restrict  
Bob's power. Particularly, when Bob tries to make a measurement in the first step
then using a standard technique we can move this measurement to the second step.}.  
Next the last qubit of $U(\QBit{\Phi_{a_0,a_1,h}})$ 
is sent to Alice\footnote{We can assume wlog that the last qubit is sent since $U$ 
is arbitrary},
she performs step 3 on these qubit and sends the classical bit $m$ back to Bob. 
Upon receiving $m$, Bob executes the second part of his attack: he performs some arbitrary measurement $\{H_{0},H_{1},H_{2},H_{3}\}$, where $H_{0}$ ($H_{1}$) 
corresponds to Bob's belief that $a_{0}=0, a_{1}=0$ (resp. $a_{0}=0, a_{1}=1$) and $H_{2}$ ($H_3$) corresponds 
to $a_{0}=1$ and $a_1=0$ (resp. $a_0=1$ and $a_1=1$). In other words, outcome corresponding to $H_{2l+k}$ implies $a_0'=l$ and $a_1'=k$.
 
The unitary transformation $U$ can be described by a set of vectors $\{V_{k}^{l,j}\}$ such that 
$U(\QBit{v}\otimes \QBit{l,j})=\QBit{V_{0}^{l,j}}\otimes \QBit{0}+\QBit{V_{1}^{l,j}}\otimes \QBit{1}$.
Or alternatively in diagonal basis, by a set of vectors $\{W_{k}^{l,j}\}$ such that 
$U(\QBit{v}\otimes \QBit{l_{\times},j_{\times}})=
\QBit{W_{0}^{l,j}}\otimes \QBit{0_{\times}}+\QBit{W_{1}^{l,j}}\otimes \QBit{1_{\times}}$.

We present now, an intuitive, brief summary of the proof. Informally, we can think of $U$ as about some kind of disturbance of the qubit $R_\alpha\QBit{a_0\oplus h}$ being sent back to Alice. First, we will show that in order to cheat Bob's $U$ has to accumulate after Step 2, till the end of the protocol, some information about the value of $a_0\oplus h$ hidden in this qubit. On the other hand, to get the proper result i.e. the value of $a_0$, this qubit's actual information about encoded value has to be disturbed at the smallest possible degree. That implies for Bob a necessity of some sort of cloning that qubit, which turns out to impose the desired bounds on possible cheating. We show this by first reducing the task of cloning to one where no additional hint in the form of $R_\alpha\QBit{a_1\oplus h}$ is provided and then an analysis of this simplified process. Therefore, the proof indicates that the hardness of cheating the protocol is contained in the necessity of cloning, which gives us a sort of quantitative non-cloning theorem. Although, it seems to concern only our particular implementation of the protocol, we believe that this scenario is useful enough to be of independent interests.

%   I.e. we show that the task of Bob's cloning an xor of two bits, say $x\oplus y$ encoded with equal probability in diagonal or perpendicular basis, in such a way that after giving it back to Alice and acquiring her measurement outcome xor-ed with $y$: mainly, get knowledge about value of $x$ and possibly: get some information about $y$. Although, it seems to concern only our particular implementation of the protocol, we believe that this scenario is useful enough to be of independent interests. 

We analyze first Bob's information gain about $a_1$. Wlog we may assume that Bob can distinguish better between two values of $a_1$ if $a_0=0$. That is 
$$
\Prob_{a_1\in_R\{0,1\}}[\mbox{$a'_1 = a_1$}|a_0=0]\geq \Prob_{a_1\in_R\{0,1\}}[\mbox{$a'_1 = a_1$}|a_0=1].
$$

Let now $\rho_{j,k,l}$ be a density matrix of the system before Bob's final measurement, 
corresponding to $\alpha=j\cdot \frac{1}{2}$, $h=k$, $a_{1}=l$ and $a_0=0$. 
The advantage $\delta$ of Bob in this case (i.e. $\delta$ such that  
$\Prob[a'_{1}=a_{1}\ |\ a_0=0] = 1/2 + \delta$) can be estimated by 
Lemma~\ref{Lemm:AB:prob:trace} by Bob's ability to distinguish between the
following density matrices:
\[
\begin{array}{ll}
 \frac{1}{4}(\rho_{0,0,0}+\rho_{1,0,0}+\rho_{0,1,0}+\rho_{1,1,0}) 
    &\quad \mbox{(case $a_1=0$), and} \\[2mm]  %%& & \label{eq:rep:z}\\
 \frac{1}{4}(\rho_{0,0,1}+\rho_{1,0,1}+\rho_{0,1,1}+\rho_{1,1,1}) 
    &\quad \mbox{(case $a_1=1$).} \\[2mm] %% & &. \label{eq:rep:o}
\end{array}
\]
Using the triangle inequality we get that for the measurement $\cO$ performed by Bob
\begin{equation} \label{eq:bound:on:delta}
  \delta \le \frac{1}{8}(|\rho^{\cO}_{0,0,0}  - \rho^{\cO}_{0,1,1}|_1 +
                         |\rho^{\cO}_{1,1,0}  - \rho^{\cO}_{1,0,1}|_1 +
                         |\rho^{\cO}_{0,1,0}  - \rho^{\cO}_{0,0,1}|_1 +
                         |\rho^{\cO}_{1,0,0}  - \rho^{\cO}_{1,1,1}|_1 ).
\end{equation}
Each component corresponds to different values of $\alpha$ and $h\oplus a_{1}$. And each component is symmetric to the other in such a way that there exists a straight-forward local transformation for Bob (i.e. appropriate rotation of the computational basis on one or both qubits) which transform any of above components onto another. So, we can assume wlog that the advantage in distinguishing between $\rho_{0,0,0}$ and $\rho_{0,1,1}$ $\delta_0=|\rho^{\cO}_{0,0,0}  - \rho^{\cO}_{0,1,1}|_1$ is 
the maximum component in the right-hand side of the inequality (\ref{eq:bound:on:delta}) and therefore we have $\delta\le \frac{1}{2}\delta_0$. Let, for short, $\gamma_0=\rho_{0,0,0}$ and $\gamma_1=\rho_{0,1,1}$. One can easily calculate that
\begin{eqnarray}
\gamma_0 & = & \mQBit{0}{0}\otimes \mQBit{V_{0}^{00}}{V_{0}^{00}}+\mQBit{1}{1}\otimes \mQBit{V_{1}^{00}}{V_{1}^{00}}\\
\gamma_1 & = & \mQBit{0}{0}\otimes \mQBit{V_{1}^{01}}{V_{1}^{01}}+\mQBit{1}{1}\otimes \mQBit{V_{0}^{01}}{V_{0}^{01}}.
\end{eqnarray}
As we can see to each value of $m$ in above density matrices corresponds a pair of vectors which are critical for Bob's cheating. I.e. the better they can be distinguishable by his measurement the greater is his advantage. %If we had for instance that $\skQBit{V_{0}^{00}}{V_{1}^{01}}=0$ which corresponds to copying of the value $a_0\oplus h$ encoded in perpendicular basis, 
 But, as we will see later, this fact introduces perturbation of the indication of the value of $a_0$.   

%The first two components correspond to distinguishing between density matrices 
%describing the same values of  and analogously the 
%third and the last component correspond to the cases when $\alpha$ and $h\oplus a_{1}$ 
%have different values. One can easily calculate that 
%
%\begin{eqnarray}
%\rho_{1,0,1}: & & \mQBit{0}{0}\otimes \mQBit{W_{0}^{10}}{W_{0}^{10}}+
%                  \mQBit{1}{1}\otimes \mQBit{W_{1}^{10}}{W_{1}^{10}}\\
%\rho_{1,1,0}: & & \mQBit{0}{0}\otimes \mQBit{W_{1}^{11}}{W_{1}^{11}}+
%                  \mQBit{1}{1}\otimes \mQBit{W_{0}^{11}}{W_{0}^{11}}
%\end{eqnarray}
%and that representations of $\rho_{1,0,0}$, $\rho_{1,1,1}$, $\rho_{0,0,1}$, $\rho_{0,1,0}$ 
%are symmetric to the representations of
%$\rho_{0,0,0}$, $\rho_{0,1,1}$, $\rho_{1,0,1}$, $\rho_{1,1,0}$ respectively,
%in such  a sense that by applying local transformation of the basis, Bob can swap the roles of $V$ and $W$.
%
%So, we can assume wlog that the advantage in distinguishing between $\rho_{0,0,0}$ and $\rho_{0,1,1}$ 
%$\delta_0=|\rho^{\cO}_{0,0,0}  - \rho^{\cO}_{0,1,1}|_1$ is 
%the maximum component in the right-hand side of the inequality (\ref{eq:bound:on:delta}) 
%and let, for short, $\gamma_0=\rho_{0,0,0}$ and $\gamma_1=\rho_{0,1,1}$. 
%Obviously, $\delta\le \frac{1}{2}\delta_0$. 
%Assume Bob's advantage in distinguishing between $\gamma_0$ and $\gamma_1$ 
%is $\delta_0$.

First, we take a look on the measurements $H_0$, $H_1$ performed by Bob.  
Let us define $\sigma_{2m+p}$ for $p,m\in\{0,1\}$ as follows
\[
  \sigma_{2m+p}=
  \left\{
  \begin{array}[c]{ll}
     |tr(H_p\mQBit{0V_{p}^{0p}}{0V_{p}^{0p}}H_p^\dagger)-
      tr(H_p\mQBit{0V_{1-p}^{0(1-p)}}
                  {0V_{1-p}^{0(1-p)}}H_p^\dagger)|
   & \mbox{if $m=0$,} \\[3mm]
     |tr(H_p\mQBit{1V_{1-p}^{0p}}{1V_{1-p}^{0p}}H_p^\dagger)-
      tr(H_p\mQBit{1V_{{p}}^{0(1-p)}}
                  {1V_{{p}}^{0(1-p)}}H_p^\dagger)|
   & \mbox{if $m=1$. }
  \end{array}\right.
\]
Let for $m=0$, $p_0\in\{0,1\}$ be such that 
$\sigma_{p_{0}}\geq \sigma_{1-p_{0}}$ and 
similarly, for $m=1$  let $p_1\in\{0,1\}$ be  such that 
$\sigma_{2+p_1}\geq \sigma_{2+(1-p_{1})}$. 
Then we get
\[
\begin{array}{lcl}
  | \gamma_0^{\cO}  -  \gamma_1^{\cO} |_1 
                            &=& \sum_{t=0}^3
                                |tr(H_t \gamma_0 H_t^\dagger) - tr(H_t \gamma_1 H_t^\dagger)| \\[2mm]
                            &\le& 2(\sigma_{p_0}+ \sigma_{2+p_{1}}) +
                                  \sum_{t=2}^3 |tr(H_t \gamma_0 H_t^\dagger) - tr(H_t \gamma_1 H_t^\dagger)|.
\end{array}
\]

We should see first that the second term in the above sum corresponds to advantage in distinguishing between two values of $a_1$ by measurement $H_2,H_3$ in case of $a_0=0$. But those subspaces reflect Bob's belief that $a_0=1$. Therefore, we have that 
$$
\sum_{t=2}^3 |tr(H_t \gamma_0 H_t^\dagger) - tr(H_t \gamma_1 H_t^\dagger)|\leq \Prob_{a_0,a_1\in_R\{0,1\}}[\mbox{$a_0' \neq a_0$}|a_0=0].
$$
 So, we can neglect this term because it is of the order of the square of the advantage (if not then our lemma would be proved).
Hence we get:
$
  \frac{\delta_0}{2}\le  \sigma_{p_0}+\sigma_{2+p_{1}}
$.

Now, we define projection $O_m$ as follows. For $m=0$
let $O_{0}$ be the normalized orthogonal projection of $\QBit{0V_{p_0}^{0p_0}}$ 
onto the subspace $H_{p_0}$ if 
\[
  tr(H_{p_0}\mQBit{0V_{p_0}^{0p_0}}{0V_{p_0}^{0p_0}}H_{p_0}^\dagger) \ \ge\
  tr(H_{p_0}\mQBit{0V_{1-p_0}^{0(1-p_0)}}{0V_{1-p_0}^{0(1-p_0)}}H_{p_0}^\dagger).
\]
Otherwise, let $O_{0}$ be the normalized orthogonal projection of 
$\QBit{0V_{1-p_0}^{0(1-p_0)}}$ onto $H_{p_0}$.
%>> This remark is false and in fact 
%>> we do not need this additional information.
% i.e. $H_{p_0}=\mQBit{O_0}{O_0}+H_{p_0}'$
%where $tr(H_{p_0}'\mQBit{0V_{p_0}^{0p_0}}{0V_{p_0}^{0p_0}}H_{p_0}')=0$.
Analogously, we define $O_{1}$ as a normalized orthogonal projection of 
$\QBit{1V_{1-p_1}^{0p_1}}$ onto the subspace $H_{p_1}$ if 
\[
  tr(H_{p_1}\mQBit{1V_{1-p_1}^{0p_1}}{1V_{1-p_1}^{0p}}H_{p_1}^\dagger) \ \ge\
  tr(H_{p_1}\mQBit{1V_{{p_1}}^{0(1-p_1)}}{1V_{{p_1}}^{0(1-p_1)}}H_{p_1}^\dagger)
\]
else $O_{1}$ is a normalized orthogonal projection of 
$\QBit{1V_{p_1}^{0(1-p_1)}}$ onto $H_{p_1}$.
%, i.e. 
%$H_{p_1}=\mQBit{1V_{\overline{p_1}}^{0p_1}}{1V_{\overline{p_1}}^{0p_1}}+H_{p_1}''$ where 
%$tr(H_{p_1}''\mQBit{1V_{\overline{p_1}}^{0p_1}}{1V_{\overline{p_1}}^{0p_1}})=0$.
Hence we get 
$$
%\begin{eqnarray}
\sigma_{p_0}\leq ||\skQBit{0V_{p_0}^{0p_0}}{O_0}|^2-|\skQBit{0V_{1-p_0}^{0(1-p_0)}}{O_0}|^2|,
\quad
\sigma_{2+p_1}\leq ||\skQBit{1V_{1-p_1}^{0p_1}}{O_1}|^2-|\skQBit{1V_{p_1}^{0(1-{p_1})}}{O_1}|^2|.
%\end{eqnarray}
$$
%
%\[
%  |\skQBit{0V_{p_0}^{0p_0}}{O_0}|^2=
%  tr(H_{p_0}\mQBit{0V_{p_0}^{0p_0}}{0V_{p_0}^{0p_0}}H_{p_0})\geq {\bf ???}
%  tr(H_{p_0}\mQBit{0V_{\overline{p_0}}^{0\overline{p_0}}}
%                  {0V_{\overline{p_0}}^{0\overline{p_0}}}H_{p_0})\geq 
%  |\skQBit{0V_{\overline{p_0}}^{0\overline{p_0}}}{O_0}|^2
%\] 

%\[
%  |\skQBit{1V_{\overline{p_1}}^{0p_1}}{O_1}|^2=
%  tr(H_{p_1}\mQBit{1V_{\overline{p_1}}^{0p_1}}{1V_{\overline{p_1}}^{0p_1}}H_{p_1})\geq {\bf ???}
%  tr(H_{p_1}\mQBit{1V_{{p_1}}^{0\overline{p_m}}}{1V_{{p_m}}^{0\overline{p_m}}}H_{p_1})\geq 
%  |\skQBit{1V_{{p_1}}^{0\overline{p_1}}}{O_1}|^2
%\] 
% 
%
%\vspace{15pt}

We would like now to investigate the probability of obtaining the correct result.  
Recall that $\Prob[a_{1}=0]=\frac{1}{2}$. We should first note that the density 
matrices corresponding to initial configuration of the second qubit $R_\alpha\QBit{a_1\oplus h}$ 
is now exactly $\frac{1}{2}\mQBit{0}{0}+\frac{1}{2}\mQBit{1}{1}$ even if we know $h$ and $\alpha$.
So, from the point of view of the protocol those two configurations are indistinguishable. 
Therefore, we can substitute the second qubit from the initial configuration with a random bit 
$r$ encoded in perpendicular basis and the probability of obtaining proper result is unchanged. 
We analyze the probability of computing the correct result in case of $r=0$.
Note, that the vectors $\{V_{k}^{0,j}\}_{k,j}$ still describe $U$, but vectors 
$\{W_{k}^{0j}\}_{k,j}$ are different, defined by $U$ acting now on initial configuration 
$\QBit{v}\otimes \QBit{0}\otimes R_\alpha\QBit{j} $, with $\alpha=\frac{1}{2}$. 
We investigate the correspondence between $\{V_{k}^{0j}\}_{k,j}$ and 
the new vectors. %$\{W_{k}^{0j}\}_{k,j}$. 
For $j=0$ we have:
\[
\begin{array}{rcl}
U(\QBit{v00_\times})
  &=&   \frac{1}{\sqrt{2}} U(\QBit{v00} - \QBit{v01}) \ = \
  \frac{1}{\sqrt{2}} (V_{0}^{00}\QBit{0}+V_{1}^{00}\QBit{1}-V_{0}^{01}\QBit{0}-V_{1}^{01}\QBit{1})\\[2mm]
 % &=&   \frac{1}{2} (V_{0}^{00}(\QBit{0_\times}+\QBit{1_\times})+
 %                    V_{1}^{00}(-\QBit{0_\times}+\QBit{1_\times})-
 %                    V_{0}^{01}(\QBit{0_\times}+\QBit{1_\times})-
 %                    V_{1}^{01}(-\QBit{0_\times}+\QBit{1_\times}))\\[2mm]
  &=&   \frac{1}{2} ((V_{0}^{00}-V_{1}^{00}-V_{0}^{01}+V_{1}^{01})\QBit{0_\times}+
                     (V_{0}^{00}+V_{1}^{00}-V_{0}^{01}-V_{1}^{01})\QBit{1_\times})).
\end{array}
\]
Similarly, for $j=1$ we have:
\[
\begin{array}{rcl}
U(\QBit{v01_\times})
  &=&   \frac{1}{\sqrt{2}} U(\QBit{v00} + \QBit{v01})\ = \
  \frac{1}{\sqrt{2}} (V_{0}^{00}\QBit{0}+V_{1}^{00}\QBit{1}+V_{0}^{01}\QBit{0}+V_{1}^{01}\QBit{1})\\[2mm]
%  &=&   \frac{1}{2} (V_{0}^{00}(\QBit{0_\times}+\QBit{1_\times})+
%                     V_{1}^{00}(-\QBit{0_\times}+\QBit{1_\times})+
%                     V_{0}^{01}(\QBit{0_\times}+\QBit{1_\times})+
%                     V_{1}^{01}(-\QBit{0_\times}+\QBit{1_\times}))\\[2mm]
  &=&   \frac{1}{2} ((V_{0}^{00}-V_{1}^{00}+V_{0}^{01}-V_{1}^{01})\QBit{0_\times}+
                     (V_{0}^{00}+V_{1}^{00}+V_{0}^{01}+V_{1}^{01})\QBit{1_\times})).
\end{array}
\]%After a little algebra 
Thus, let us denote these vectors by  %we get that:let us denote the 
\begin{eqnarray*}
\widetilde{W}_{0}^{00} = \frac{1}{2}((V_{0}^{00}+V_{1}^{01})-(V_{0}^{01}+V_{1}^{00})), & &
\widetilde{W}_{1}^{00} = \frac{1}{2}((V_{0}^{00}-V_{1}^{01})-(V_{0}^{01}-V_{1}^{00})),\\
\widetilde{W}_{0}^{01} = \frac{1}{2}((V_{0}^{00}-V_{1}^{01})+(V_{0}^{01}-V_{1}^{00})), & &
\widetilde{W}_{1}^{01} = \frac{1}{2}((V_{0}^{00}+V_{1}^{01})+(V_{0}^{01}+V_{1}^{00})).
\end{eqnarray*}

In order to obtain the correct result Bob has to distinguish between the density matrices corresponding to two values of $a_{0}$. In particular, he has to distinguish between density matrices $\gamma_0'$, $\gamma_1'$ corresponding to two possible values of $a_{0}$ knowing that $m=0$. These density matrices are: 
\begin{eqnarray}
        \gamma_0'&=& \frac{1}{4}\mQBit{0}{0}\otimes(\mQBit{V_{0}^{00}}{V_{0}^{00}}+\mQBit{V_{1}^{01}}{V_{1}^{01}}+\mQBit{\widetilde{W}_{0}^{00}}{\widetilde{W}_{0}^{00}}+\mQBit{\widetilde{W}_{1}^{01}}{\widetilde{W}_{1}^{01}}),\\     
        \gamma_1'&=& \frac{1}{4}\mQBit{0}{0}\otimes(\mQBit{V_{0}^{01}}{V_{0}^{01}}+\mQBit{V_{1}^{00}}{V_{1}^{00}}+\mQBit{\widetilde{W}_{0}^{01}}{\widetilde{W}_{0}^{01}}+\mQBit{\widetilde{W}_{1}^{00}}{\widetilde{W}_{1}^{00}}).
\end{eqnarray}
Now, the probability of failure i.e. the probability that in case of $m=0$ Bob's measurement indicates that $a_{0}=0$ if in fact it is $a_0=1$, is at least 
$$
tr(H_{p_0}\gamma_1'H_{p_0}^\dagger)\geq tr(\mQBit{O_{0}}{O_0}\gamma_1')
=\frac{1}{4}(|\skQBit{0V_{0}^{01}}{O_0}|^2+|\skQBit{0V_{1}^{00}}{O_0}|^2+|\skQBit{0\widetilde{W}_{0}^{01}}{O_0}|^2+|\skQBit{0\widetilde{W}_{1}^{00}}{O_0}|^2).
$$
But since the fact that
$$
\widetilde{W}_{0}^{01}=\frac{1}{2}((V_{0}^{00}-V_{1}^{01})+(V_{0}^{01}-V_{1}^{00})),\qquad  
\widetilde{W}_{1}^{00}=\frac{1}{2}((V_{0}^{00}-V_{1}^{01})-(V_{0}^{01}-V_{1}^{00})),
$$
and the parallelogram law ($|a+b|^2+|a-b|^2=2|a|^2+2|b|^2$), we have that this probability is at least

\begin{center} 
\begin{tabular}{c}
$\frac{1}{4}(|\skQBit{0\widetilde{W}_{0}^{01}}{O_0}|^2+|\skQBit{0\widetilde{W}_{1}^{00}}{O_0}|^2)\geq \frac{1}{8}|\skQBit{0V_{0}^{00}}{O_0}-\skQBit{0V_{1}^{01}}{O_0}|^2$ \\[2mm]
$\geq \frac{1}{32}(|\skQBit{0V_{0}^{00}}{O_0}|-|\skQBit{0V_{1}^{01}}{O_0}|)^2 (|\skQBit{0V_{0}^{00}}{O_0}|+|\skQBit{0V_{1}^{01}}{O_0}|)^2$ \\[2mm]
$\geq\frac{1}{32}(|\skQBit{0V_{0}^{00}}{O_0}|^2-|\skQBit{0V_{1}^{01}}{O_0}|^2)^2 \geq \frac{\sigma_{p_0}^2}{32}.$\\ 
\end{tabular}
\end{center}

Similarly we analyze density matrices $\gamma_0''$, $\gamma_1''$ corresponding to two possible values of $a_{0}$ knowing that $m=1$. These density matrices are equal to resp. $\gamma_1'$ and $\gamma_0'$ after changing $\mQBit{0}{0}$ to $\mQBit{1}{1}$.
Now, by repeating completely analogous estimation of failure's probability with usage of vectors $\QBit{V_{0}^{01}}$, $\QBit{V_{1}^{00}}$, $\QBit{\widetilde{W}_{0}^{00}}$ and $\QBit{\widetilde{W}_{1}^{01}}$, we get that this probability is at least $\frac{\sigma_{2+p_1}^2}{32}$. Therefore, since the vectors involved in imposing failure in both cases are distinct, we conclude that 
%\begin{center}
$\Prob_{a_1\in_R\{0,1\}}[\mbox{$a'_0 \neq a_0$}|r=0]\geq \frac{\sigma_{p_0}^2+\sigma_{2+p_1}^2}{32}$.
%\end{center}
%
Hence we have 
\begin{center}
 \begin{tabular}{c}
$\Prob_{a_1\in_R\{0,1\}}[\mbox{$a'_0 \neq a_0$}]=\frac{1}{2}\Prob_{a_1\in_R\{0,1\}}[\mbox{$a'_0 \neq a_0$}|r=0]+\frac{1}{2}\Prob_{a_1\in_R\{0,1\}}[\mbox{$a'_0 \neq a_0$}|r=1]$\\
$\geq \frac{\sigma_{p_0}^2+\sigma_{2+p_1}^2}{64}\geq \frac{\delta^2}{128}$\\
\end{tabular}
\end{center}
 and the lemma is proved.

%$$(\frac{1}{2}(2(\sigma_{p_{0}}+\sigma_{2+p_1})+2(\sigma_{p_{0}'}'+\sigma_{2+p_1'}')),\frac{1}{64}(\sigma_{p_{0}}^2+\sigma_{p_{1}}^2+\sigma_{p_0'}^{'2}+\sigma_{p_1'}^{'2}))$-security.Where by primed quantities we mean those having the same definition as unprimed ones but corresponding to case of $r=1$. This yields in worst case $(16\epsilon,\epsilon^2)$-security. 

Finally, it is worth mentioning that the value of $m$ doesn't need to be correlated in any way with value of $a_{i}$. That is, Bob by using entanglement (for instance, straightforward use of Bell states) can make the value of $m$ independent of $a_{i}$ and still acquire perfect knowledge about $a_{i}$. He uses simple error-correction to know whether $m=a_{i}$ or $m=1-a_{i}$. His problems with determining whether flip has occurred, start only when he wants additionally to accumulate some information about the value of $a_{i}\oplus h$. 
\end{Proof}

To see that this quadratical bound can be achieved consider the following cheating strategy. 
Let $U^{\ast}$ be such that $U^{\ast}(\QBit{v}\otimes\QBit{l,j})=\QBit{v_j}\otimes\QBit{l,j}$. So, $\QBit{V_{j}^{l,j}}=\QBit{v_j}\otimes \QBit{l}$ and $\QBit{V_{1-j}^{l,j}}=0$. Moreover, let $\skQBit{v_0}{v_{1}}=\sqrt{1-\varepsilon}$. As we can see, usage of $U^{\ast}$ accumulates some information about value of $j=a_0\oplus h$ by marking it with two non-parallel (therefore possible to distinguish) vectors in Bob's system.
We do now the following. We use $U^{\ast}$ on $\QBit{v}\otimes R_\alpha\QBit{a_1\oplus h}\oplus R_\alpha\QBit{a_0\oplus h}$ and send the last qubit to Alice.
When we get the message $m$ which is exactly $a_{0}$ with probability\footnote{This can be easily computed - the perturbation arises when $\alpha=\frac{1}{2}$.} of order $1-\varepsilon$, we make an optimal measurement to distinguish between $v_{0}$ and $v_{1}$. By Theorem \ref{Th:AKN} this optimal measurement has advantage of order $\sqrt{\varepsilon}$. So, after getting the outcome $j'$, we know that $\Prob[\mbox{$j' = a_{0}\oplus h$}]\geq \frac{1}{2}+\Omega(\sqrt{\varepsilon})$ and we can simply compute the value of $h'=m\oplus j'$. Having such knowledge about the value of $h'$ we can distinguish between values of $a_{1}$ encoded in the second qubit $R_{\alpha}\QBit{a_1\oplus h}$ with the advantage proportional to $\Omega(\sqrt{\varepsilon})$.

\section{Cheat Sensitive Quantum Bit Commitment} \label{sec:qot}

We recall first a formal definition of the binding and sealing property
of a quantum bit commitment. We follow here the definition by Aharonov 
et~al. \cite{ATVY00}.
Let us start with the binding property.
Assume Alice follows the bit commitment protocol and Bob is arbitrarily.
During the depositing phase Bob and Alice compute in some rounds a super-position 
$\QBit{\psi_{AB}}$  with two quantum registers: one keeping by Bob and one by Alice.
After a communication phase Bob either uses a strategy trying to convince 
Alice to $0$ or a strategy to convince Alice to  $1$.
Depending on the results of the computations Alice decides to one the values $v_B\in \{0,1,err\}$;
In case $v_B=err$ he rejects the protocol.
Let $p_i$ be the probability that Alice decides $v_B=i$,
and  $p_{err}$ be the probability that Alice decides $v_B=err$,
when Bob uses strategy $0$. Analogously, denote the probabilities 
$q_0,q_1,q_{err}$ for Bob's strategy $1$.
A protocol is $(\delta, \varepsilon)$-{\em binding} if whenever 
Alice is hones, for any Bob's strategy it is true:
if\ $p_{err},q_{err}\le \varepsilon$\ then\ $|p_0-q_0|, |p_1-q_1| \le \delta$.
A bit commitment protocol is $(\delta,\varepsilon)$-{\em sealing},
if whenever Bob is honest and deposits a bit $b$
s.t. $\Prob[b=0]=1/2$, for any Alice's strategy and a value $c$ Alice learns, 
it holds that: 
if\ $\Prob_{b\in_R \{0,1\}}[\mbox{Bob detects error}]\le \varepsilon$
       \ then\  $\Prob_{b\in_R \{0,1\}}[c=b]\le 1/2 + \delta$.
The probability is taken over $b$ taken uniformly from $\{0,1\}$
and the protocol.

\begin{Theorem}\label{Th:chet:sensit:QBC}
Using Protocol \ref{Protocol:OT} as a black-box for computing OT,
Protocol \ref{Protocol:weakQBC} is an $(4\sqrt{\varepsilon},\varepsilon)$-sealing.
Moreover, there exists a constant $\lambda>0$ such that for all strategies 
Bob uses it holds $\max\{p_{err}, q_{err}\} > \lambda$, where 
$p_{err}$ ($q_{err}$) denotes the probability that Alice decides error
when Bob uses strategy for $0$ ($1$ resp.).
\end{Theorem}
\begin{Proofsketch}
First, we note that in both calls to the OT function the inputs that come 
into play in this executions are completely uncorrelated from the point of 
view of both Alice and Bob. So, we can analyze them distinctly.

To see that this protocol is sealing we note that Alice in each call to OT 
function has to take into account that with probability $\frac{1}{2}$ Bob will 
check whether she knows what actually he has received during execution of 
this protocol. Moreover her cheating is effective only if it is not checked, 
so only with probability of $\frac{1}{2}$. By Lemma \ref{l:mal:Al}, 
if a strategy allows her to distinguish between possible values of $b'$ 
with advantage greater than $4\sqrt[2]{2\varepsilon}$ then 
$\Prob_{b'\in_R\{0,1\}}[ v_c\neq OT((a_{2c}, a_{2c+1}), b')]\ge \varepsilon$. 

In case of binding, we first notice that it is only useful for Bob to cheat in 
some particular OT execution, chosen previously by Bob, which is used in the revealing 
phase for the binding test. So wlog assume Bob cheats in the second OT execution
and that in the last step of the depositing stage he reveals $c=0$.
%Assume now that Bob gains some knowledge about $a_3, a_4$ and that 
%the probability that he predicts $a_3$ correctly is greater or equal to 
%the probability of a correct prediction for $a_4$. 
Let $a'_3$, $a'_4$, resp. 
denote the predicted values. Using the notation given in the definition of 
the binding property we get that 
$p_{err} = \Prob[a'_3\not = a_3]$, $p_0 = \Prob[a'_3 = a_3]$, and  $p_1 = 0$.
Similarly we have $q_{err} =  \Prob[a_4'\not=a_4]$, $q_0=0$, and $q_1 = \Prob[a'_4=a_4]$.
Now by Lemma~\ref{lem:mal:Bob} we get that 
if $\Prob[\mbox{$a'_i \ne a_i$}]\le \varepsilon^2$ then  
$\Prob[a'_{1-i}\ne a_{1-i}]\ge 1/2 - 16^2\varepsilon$
and for some constant $\lambda>0$ 
it follows that $\max\{ \Prob[\mbox{$a'_i \ne a_i$}], \Prob[a'_{1-i}\ne a_{1-i}] \} > \lambda$.
%resp.
%$\Prob[a'_{1-i} = a_{1-i}]\le 1/2 + 32\varepsilon$.
%Hence, if $q_{err}\le \varepsilon^2$ and $p_{err}\le \varepsilon^2$
%then we have 
%$1-\varepsilon^2 \le |p_0-q_0|, |p_1-q_1| \le 1/2 + 32 \varepsilon$
%as well as
%$$
%\frac{1}{2}-32\varepsilon \ \ \le \ \ p_{err},\ q_{err} \ \ \le \ \ \varepsilon^2\ .
%$$
%Note that this inequality can only be fulfilled if 
%$\varepsilon\ge \sqrt{\frac{513}{2}}-16\ge 0,0156$.
%We get $p_{err},\ q_{err}\ge 0,00024$.
%Hence for all $\varepsilon$ %\le 0.0156
%bounded by some strictly positive constant $s$ we get that $p_{err}$ and $q_{err}$
%cannot be both bounded by $\varepsilon^2$. Moreover, for all $\varepsilon>s$
%we have that if $p_{err},q_{err}\le \varepsilon^2$ then 
%$|p_0-q_0|, |p_1-q_1| \le 1/2 + 32 \varepsilon \le O(\varepsilon)$.
\end{Proofsketch}

\section{Concluding Remark}

In this paper a weak variant of quantum  bit commitment
is investigated. We give quantum  bit commitment scheme that is simultaneously
binding and sealing and we show that if a malicious Alice gains some information 
about the committed bit $b$ then Bob detects this with a probability 
$\Omega(\varepsilon^2)$. When Bob cheats then Alice's probability of detecting 
the cheating is greater than a constant $\lambda>0$. Using our bounds 
we get that the value is very small and an interesting task would be to improve 
the constant.

%To construct our quantum  bit commitment scheme we use 
%a protocol for $\binom{2}{1}$-OT such that if a 
%malicious party gains some knowledge then the party 
%fail in computing the correct result. An interesting problem 
%would be to construct a cheat sensitive protocol for 
%$\binom{2}{1}$-OT, i.e. such a scheme that the cheater 
%can be detected with a good probability.

\end{document}